\documentclass[preprint,10pt]{elsarticle}

\usepackage{epsfig}
\usepackage{amssymb}
\usepackage{graphicx,psfrag}
\usepackage{bm}

\newcommand{\be}{\begin{equation}}
\newcommand{\ee}{\end{equation}}
\newcommand{\bea}{\begin{eqnarray}}
\newcommand{\eea}{\end{eqnarray}}
\newcommand{\hf} {\frac{1}{2}}

\def\eq#1{(\ref{#1})}
\def\ord#1{{\cal O}(#1)}

\def\mr#1{{\mathrm{#1}}}
\def\t{\tilde}
\def\S{\mathcal{S}}
\begin{document}
\begin{frontmatter}

\title{Quantum censorship in two dimensions}

\author{V. Pangon}
\address{Frankfurt Institute for Advanced Studies, Universit\"at Frankfurt, D-60438 Frankfurt am Main,
Germany\\ Gesellschaft f\"ur Schwerionenforschung mbH, Planckstr. 1, D-64291 Darmstadt, Germany}

\author{S. Nagy}
\address{Department of Theoretical Physics, University of Debrecen, Debrecen, Hungary}

\author{J. Polonyi}
\address{Strasbourg University, CNRS-IPHC, BP28 67037 Strasbourg Cedex 2, France} 

\author{K. Sailer}
\address{Department of Theoretical Physics, University of Debrecen, Debrecen, Hungary}

\begin{abstract}
It is pointed out that increasingly attractive interactions, represented by partially concave 
local potential in the Lagrangian, may lead to the degeneracy of the blocked, renormalized
action at the gliding cutoff scale by tree-level renormalization. A quantum
counterpart of this mechanism is presented in the two-dimensional sine-Gordon model.
The presence of Quantum Censorship is conjectured which makes the loop contributions 
pile up during the renormalization and thereby realize an approximate semiclassical effect.
\end{abstract}
\end{frontmatter}

\section{Introduction} 
A necessary condition for classical physics to emerge from quantum fluctuations is 
the high excitation level density or in other words, the strong degeneracy. The first 
condition is well known from the early days of Quantum Mechanics, the second way of 
expressing it implies decoherence \cite{zeh,zurek}. This is a sharp contrast to the
usual scenario of the semiclassical limit where the trajectories around the classical one
are supposed to dominate the path integral. The solution of this apparent conflict arises from the
remark that the dominance of the path integral by the saddle point
is easy to establish for the transition amplitudes between pure states 
in a finite system only. Decoherence arises from degeneracy in a clear manner as soon 
as mixed states and density matrices are considered \cite{ed}. Another source of 
the degeneracy is the thermodynamical limit, the inclusion of soft modes in the system.
The degeneracy of a realistic system is supposed to occur in both ways.

A large degree of degeneracy opens new problems owing to the increased quantum 
fluctuations. In fact, when the integrand in the path integral becomes approximately constant
for infinitely many modes then the system is strongly coupled and renders our analytical
methods inefficient. One may wonder whether one encouters such a scenario 
in the usual path integral, corresponding to either Euclidean partition function or 
real time transition amplitude in the thermodynamical limit. The goal of this work is 
to monitor the degree of degeneracy by means of the functional RG method in a model where 
a condensate is expected .

\section{Degeneracy and functional RG method}
There have been a number of works using the functional RG method for models with 
spontaneously broken symmetry. For instance, the $O(N)$ symmetric scalar model was 
considered in Ref. \cite{ringwald} in $d$ dimensions for $N\ge3$ where the average action,
calculated in the one-loop approximation was found to be
degenerate for small enough field amplitudes. But the degeneracy appears in the 
constrained path integral, used in this scheme as a strongly coupled dimensional model 
defined on the $d-1$ dimensional sphere in the momentum space and renders the results unreliable.

The local potential was obtained for the same model in the effective action by an
approximate solution of the evolution equation \cite{tetradis}. The degeneracy has been
identified here, too, in the unstable region. The omitted terms in the approximate expression
for the potential were proportional to $\epsilon_{deg}$, the extent of the non-degeneracy of the 
action in the unstable region. The problem with this result is that the small parameter
of the loop expansion, used to derive the evolution equation of the average potential
is $\ord{\epsilon_{deg}^{-1}}$ rendering the uncertainty of the solution $\ord{\epsilon_{deg}^0}$.
The ansatz for the effective action has been improved in Ref \cite{wett}  by 
including the wave function renormalization constant but the local potential was 
reduced to a simple quartic polynomial and such a truncation prevents us to address
the issue of degeneracy. The limit of a large number of fields offers exact
solutions and the effective action was shown to be degenerate in this limit \cite{berges}.
But as soon as the inverse of the number of fields becomes comparable with some
dimensionless measure of degeneracy then this limit becomes misleading from
the point of view of our problem. In addition, the Higgs mode was totally ignored 
in favor of the Goldston modes in the argument, an unreliable approximation when 
a possible deviation from an infinitely degenerate situation is sought.

Another attempt to see the impact of the degeneracy on the dynamic was based on the
use of sharp cut-off. The tree-level evolution of the local potential of the 
blocked bare action was followed in Ref. \cite{Alexandre}. The saddle point for
the blocking when the cut-off is lowered as $k\to k-\Delta k$ is trivial for large enough $k$.
For smaller $k$ the action develops non-trivial minima for fluctuations at the gliding 
cut-off and inhomogeneous saddle points appear. The result is a dynamical Maxwell-cut, 
the degeneracy of the bare action for small wave numbers and small values of the condensate.
This method has two weak points. One is that one can not include wave function
renormalization constant when sharp cut-off is used. The other is that the variation of the bare 
action for the first appearance of the instability is $\epsilon_{deg}=\ord{\Delta k/k}$ and
the small parameter of the expansion where the loop-integral is carried out on a shell of 
thickness $\Delta k$ in the momentum space is $\ord{\Delta k/k}/\epsilon_{deg}=\ord{(\Delta k/k)^0}$.

When dealing with the average or the bare action we have to rely on the loop-expansion
eventhough the approximate or exact solution of the evolution equation resums 
partially or completely the higher loop contributions. We have seen that the degeneracy 
arising with the condensate renders these approaches unreliable. No loop expansion is evoked 
in deriving the evolution equation of the effective action in Refs. \cite{diff,berges,Polonyi_lect}
and one would hope that this method remains applicable for condensates. But the truncation
of the evolution equation brings the problem back. The effective action is convex
by definition and the resulting Maxwell-cut hides the dominant physics, related 
to domain walls as soft, non-perturbative modes behind a degenerate effective action 
functional. The evolution is described by an autonomous equation thus the information about
the dynamics of the unstable, degenerate region must be provided by the way it is joined
to the stable, non-degenerate part of the dynamics. The non-analyticity prevents us
to use the traditional Landau-Ginzburg double expansion strategy and we have no guidance left to perform
the unavoidable truncation of the functional differential equation in this region. The
generator functional, or the effective action in particular is always used in Quantum 
Field Theory as a book-keeping device to manipulate hierarchical equations for different Green-functions.
No convergence or existence issues are raised in this manner. But as soon as the functional
is born by solving non-linear equations rather than assuming the availability of functional
Taylor series we face a radically more difficult mathematical issue, reminiscent of the
bound state problem. 

There is another problem when the evolution of the effective action is considered
in the presence of condensate. The effective action approaches a singular limit,
set by the Maxwell-cut in a manner which is determined by the suppression
term introduced in this scheme. This latter stabilizes the dynamics and provides the 
analytical continuation to derive an evolution equation in the vicinity of a singularity. 
It seems that the singularity is avoided for sufficiently strongly scale dependent
suppressions \cite{litim} indicating that the analytical continuation provided by
the arbitrary suppression term may not be unique and artificial fixed points
might be generated at the edge of the non-convex region. This problem is absent for sharp cut-off 
where each mode is taken into account with its original dynamics and no analytical 
continuation is made.

We make a step forward in this work by increasing the numerical flexibility of the 
functionals followed in the evoluton for the two dimensional sine-Gordon model. We still 
rely on the gradient expansion but renounce the expansion in the field amplitude and allow 
arbitrary field dependence during the evolution in the hope that such a richer ansatz 
allows us to reach better the formation of the condensate. We shall consider the evolution 
of the bare action given by the Wegner-Houghton equation during the lowering of a sharp cut-off. 
It is found for certain values of the coupling constants that the quantum fluctuations prevent the system
from developing a degeneracy despite the emergence of the Maxwell-cut in the effective potential. 
What is new in this result is that an almost degenerate, regular dynamics is now established without 
relying on the non-unique, regulating effects of an arbitrary suppression mechanism.
Guided by an analogous problem in General Relativity such a quantum fluctuation 
generated way to reproduce the salient feature of the semiclassical physics without 
saddle points might be called the Quantum Censorship (QC) in Quantum Field Theory \cite{Pangon}.
QC may not be established in some other part of the coupling constant space because one 
can not discover a degeneracy by means of a numerical method of finite accuracy. Analytical 
methods are even less useful for such a degenerate dynamics. Therefore, the reliable clarification
of the presence or absence of QC remains to be an open problem.

\section{Saddle points of the RG equations}
We start by inspecting the way the renormalization group method, the most promising
nonperturbative tool in quantum field theory, indicates the presence of condensate
in the vacuum. The question one wants to clarify is the scale where the condensate
shows up first as a singularity in the renormalized dynamics as the cutoff of the 
theory is lowered.

The renormalized trajectory of a quantum system maps out the scale dependence 
of the effective parameters of the system. It is one of the basic tenets of the renormalization group
procedure that the critical behaviors, the singular dependence of the IR observables on
the UV parameters, builds up by scanning through infinitely wide scale regions rather by
a singularity observed at a finite scale. Correspondingly, the renormalized trajectory
should at least be continuous in the scale parameter, an expectation which has already been justified
for local quantum field theories \cite{cont}. But the continuity of the 
renormalized trajectory in the cutoff does not exclude other singularities. 

Let us consider an Euclidean theory characterized
by the action $S_k[\phi]$, $k$ being the sharp UV cutoff, and write the field variable as
$\phi+\phi'$ where the supports of $\phi$ and $\phi'$ in Fourier space are $|p|<k-\Delta k$
and $k-\Delta k<|p|<k$, respectively. An infinitesimal blocking step corresponds to 
integrating out the modes close to the cutoff, giving the functional integration
\be\label{block}
e^{-\frac1\hbar S_{k-\Delta k}[\phi]}=\int D[\phi']e^{-\frac1\hbar S_k[\phi+\phi']},
\ee
which may possess a saddle point. The derivatives of the trajectory with respect to the
cutoff are obviously singular at the scale where this saddle point appears or disappears. 
Such a tree-level renormalization
has been found in the spinodal instability (SI) regions and was responsible for the degeneracy of 
the blocked action at the cutoff scale for certain homogeneous background field $\phi$ 
in the $\phi^4$ and the sine-Gordon (SG) model \cite{Alexandre,Nandori_SG}, a dynamical generalization 
of the Maxwell-cut. 

The blocking \eq{block} yields a functional finite difference equation whose
solution lies well beyond our analytical capabilities. 
It is usually handled by imposing rather simple restrictions,
either by ignoring altogether the loop contributions to the blocking
or restricting the evolution of the action into few coupling constants. In the context
of the SG model one retains some Fourier coefficients of $V_k(\phi)$ in the local potential approximation \cite{Nandori_SG},
\be\label{action}
S_k[\phi]=\int_x\left[\hf(\partial\phi_x)^2+V_k(\phi_x)\right].
\ee
The saddle points considered were plane waves
and the degeneracy of the action for modes at the cutoff was found by recovering the 
potential $V^{SI}_k(\phi)=-k^2\phi^2/2$  in the SI region, in a certain interval for 
$\phi$. The inhomogeneous saddle points generate non-perturbative soft modes,
the zero modes corresponding to the breakdown of the external, space-time symmetries. 
Beyond these intervals for the field the theory
appeared to be stable, without unexpected soft modes. At the end of the intervals, at the
border of the stable and unstable regions the loop corrections make the potential 
non-analytical \cite{Alexandre} and the truncation of the potential, behind any expansion 
scheme, becomes highly suspicious.

Beyond the problem of justifying the omission of the loop correction or the possible 
non-analytical structures in the potential there is an even more fundamental issue here.
The finite difference equation \eq{block} contains infinitely
many higher loop contributions which are suppressed by the small parameter
$\epsilon_k=\hbar\Delta k/k|\ln\lambda_\mr{min}|$, $\lambda_\mr{min}$ 
being the smallest eigenvalue of $\delta^2S[\phi]/\delta\phi\delta\phi$.  
The availability of the loop expansion is assumed in deriving the evolution 
equation which in turn resums the expansion in the differential equation limit, 
$\Delta k/k\to0$. There are two ways $\epsilon_k$ can become large,
either for $k\to0$ or for $\lambda_\mr{min}\to0$. The first possibility,
a singular thermodynamical limit, is discarded for the usual, local models.
The second alternative, the case of degenerate action is more realistic.
Once the blocked action becomes exactly degenerate at the cutoff scale where
it is supposed to describe best the dynamics then the 
integral in \eq{block} obeys no expansion anymore and we have
no analytical tool left to tackle the problem. Therefore, certain singularities
of the renormalized trajectory, such as the degeneracy of the action may have serious
consequences.

\section{SG model} 
The theory considered in this work is defined by the action of \eq{action}
in two-dimensional Euclidean space-time where the bare potential at the initial cutoff,
$k_\mr{init}=\Lambda=1$, is $V_B(\Phi)=k^2\t u_B\cos(\sqrt{8\pi}\beta_r\Phi)$. 
It exhibits a $Z_2$ symmetry
$\phi(x)\to-\phi(x)$ and periodicity in the internal space, $\phi(x)\to\phi(x)+2\pi/\beta$
 $(\beta= \sqrt{8\pi}\beta_r)$.
The evolution of the local potential is governed by the Wegner--Houghton equation 
\cite{Wegner} in $d=2$
\be
(2+k\partial_k) \t V_k(\phi) = -\frac1{4\pi}\ln\left(1+ \t V''_k(\phi)\right),
\label{wh}
\ee
in terms of the dimensionless potential $\t V_k = k^{-2} V_k$, in the absence of the 
non-trivial saddle point in \eq{block}. Note that the argument of the logarithm function is the
restoring force acting on the quantum fluctuations at the cutoff and should be 
positive to justify the loop expansion. When the argument becomes negative then this equation is not
valid anymore and the saddle point contributions to the integral of \eq{block} have to be taken in account. 

The model is known to exhibit two phases \cite{Coleman_bos,sine-G,Nandori_SG,Nagy_SG},
separated by the Coleman point, $\beta_r=1$. The simplest indication is the change of the 
sign of the one-loop beta-function for $\t u$ at this point, 
$\t u_k\approx\t u_B(k^2/\Lambda^2)^{\beta_r^2-1}$. The phase $\beta_r>1$ or $\beta_r<1$
preserves or breaks the internal symmetries, respectively. The phase with broken symmetry
is equivalent with the neutral sector of the massive Thirring model
\cite{Coleman_bos} and the neutral Coulomb-gas \cite{Samuel}.
The lattice regulated SG model can be mapped to the planar XY model \cite{Huang}
providing a non-perturbative RG flow.

The symmetry broken phase contains further special points. Higher, $n$-th order perturbative
contributions generate the potential $u_n\cos(\sqrt{8\pi}\beta_rn\Phi)$ and the corresponding
coupling strength is renormalizable or non-renormalizable for $\beta_r<\beta_r^{(n)}=1/n$ or
$\beta_r>\beta_r^{(n)}$, respectively. Therefore, the theory with $\beta_r^{(n+1)}<\beta_r\le \beta_r^{(n)}$
has $n$ perturbatively renormalizable parameters apart of $\beta$.
The UV scaling regime was found to be very limited due to  intermediate scaling laws, 
appearing in between the UV and the approximate SI scaling regimes where all coupling 
constants grow with decreasing  cutoff \cite{Nagy_SG}.
The point $\beta_r^{(4)}$ deserves special attention. 
The duality established in the Villain model \cite{Jose}, $(\beta,u,z)\to(2\pi/\beta,2z,u/2)$
where $z$ denotes the vortex fugacity, maps the XY model without external field, $u=0$, into 
the continuum SG model \cite{Huang}.
A special feature of the continuum SG model is its non-periodic kinetic energy which 
suppresses the vortices in the XY model context, configurations with point singularity. 
Thus the $z=0$ plane corresponds to the continuum SG model, studied in this work.  
The dual of the Coleman-point $\beta_r=1$ is the Kosterlitz-Thouless critical point, $\beta_r^{(4)}$.

The SG model is similar to non-Abelian gauge theories in what the 
field variable is compact. The effective potential of a compact variable
is flat, the constant being the only function which is periodic and convex in the same time.
Therefore, the effective potential, $V_{k=0}(\Phi)$ can not distinguish the phases
and one expects similar phenomena in non-Abelian gauge theories, as well.
It was found that the potential $\t V_k(\Phi)$, expressed in units of the cutoff,
solves this problem and can be used to identify the phase structure \cite{Nandori_SG}.

It is natural to represent the periodic potential of the SG model by a Fourier series.
But the Fourier series of the potential $V^{SI}_k(\phi)=-k^2\phi^2/2$ for
$-\pi/\beta<\phi<\pi/\beta$ with its periodic extension seen approximately in the IR 
region by following the evolution of a truncated Fourier series \cite{Nagy_SG}, 
converges badly and all we can ascertain is that an approximate degeneracy occurs in the
symmetry broken phase. 
Note that the Fourier-expansion based numerical solution of different evolution equations
\cite{Polchinski,diff,Polonyi_lect,Pawlowski} suffers the same problem 
and it is difficult to decide whether the SI occurs or not.

We avoid the limitations of a truncated series by solving the evolution equation for
unconstrained potential numerically. The potential is represented in the algorithm by 
a spline i.e. a piecewise Chebyshev polynomial \cite{Pangon}. The evolution of the coefficients of the Chebyshev 
polynomials are followed in this method and the linear algebra employed becomes singular
for degenerate actions. The internal consistency checks of the algorithm,
controlling the derivatives and the integration adjust the step size, $\Delta k$
dynamically and stops the execution of the computer program
when the a partial error in the algorithm reaches the precision of the number representation
in the computer.

\section{Coleman point} 
The theory with $\beta_r=1$ separates two phases \cite{Coleman_bos}.
The RG flow in the $\beta_r>1$ phase which is usually referred to as the non-renormalizable 
phase gives a simple evolution due to the smallness of $\t u$. What is
more interesting for us is that the potential barrier between two neighboring
minima is thinner for large $\beta_r$ and the fluctuations can "fill up" 
the minima easier. The result is QC, the stable, 
loop-generated and gradual approach of the potential $V^{SI}_k(\phi)$ as the cutoff 
is lowered with the establishment of exactly degenerate action for $k\to0$ only.
This scenario was actually established in this phase by following the evolution of the local potential,
represented by a truncated Fourier expansion \cite{Nagy_SG,Nagy_MSG}.
The small $\beta_r$, or renormalizable phase shows an interesting, more
involved structure because the barrier between the minima of the potential is wider
and QC is more difficult to realize. The approximation \cite{Nagy_SG,Nagy_MSG}
indeed leads here to degenerate action and to tree-level renormalization and
QC is prevented to act. The result is a super-universal,  strictly bare parameter
independent shape for the local potential within the unstable region.
\begin{figure}[ht]
\begin{center}
\epsfig{file=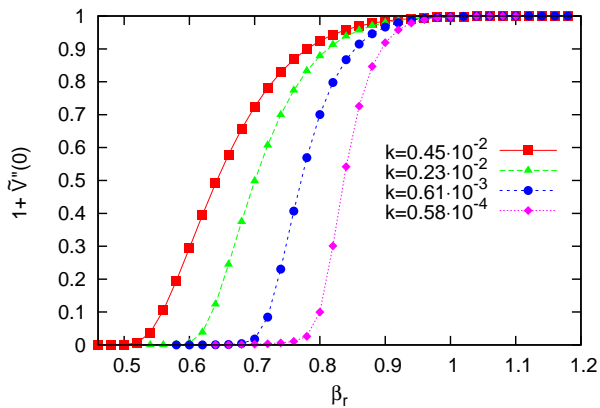,width=5.3cm,angle=-90}
\caption{\label{fig:coleman}
The second derivative of the action is plotted as the function of the
relative parameter $\beta_r$, for various values of the scale $k$.
It goes to approximately zero when $\beta_r < 1$ and to 1 when $\beta_r > 1$.
} 
\end{center}
\end{figure}

The numerical solution of the evolution equation with unconstrained potential
confirms the flattening of the dimensionless potential for $\beta_r>1$, as one 
can see in Fig. \ref{fig:coleman} where $1+\t V''(\phi=0)$ is plotted for different 
values of the cutoff $k$. What we see in the symmetry broken phase, $\beta_r<1$,
is that the action is nearly degenerate, $1+\t V''(0)\approx0$, the argument of 
the logarithm function in \eq{wh} is nearly vanishing and the evolution
is very close to be singular.

\section{Descent in $\beta$} 
What is the fate of QC in the small $\beta_r$ phase? As mentioned above, neither analytical nor numerical methods 
are available to our knowledge to answer this question. No analytical method is known 
to regulate and to handle functional integrals with constant integrands. Even if 
we grant the evolution equation \eq{wh}, no numerical algorithm, realized by 
finite computing power can distinguish exact degeneracy from small but finite 
variation. What is left is to collect circumstantial evidences to support 
our conjecture, spelled out below.

\begin{figure}[ht] 
\begin{center}
\epsfig{file=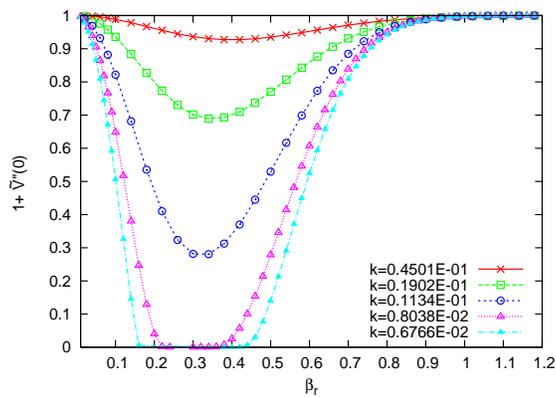,width=5.3cm,angle=-90}
\caption{\label{fig:phase}
The curvature in $\phi=0$ of the action as the function of $\beta_r$ for various scales. 
It displays maximal speed for the flow at $\beta_r\sim\beta_r^{(3)}$.} 
\end{center} 
\end{figure}

We start on the analytical side, by noting that the almost degeneracy of the 
action at the cutoff as shown in Fig. \ref{fig:coleman} generates large amplitude, 
non-perturbative modes with small but finite wave numbers, the hallmark of 
spontaneous symmetry breaking. It is the fundamental group symmetry
which might break at this point, the invariance of the theory with respect 
to the shift $\phi(x)\to\phi(x)+2\pi/\beta$. The breakdown of this symmetry is 
realized by restricting the functional integral over field configurations
with a given number of topological charge. In fact, the derivation of the correct Schwinger-Dyson
hierarchical equations, obtained formally by performing infinitesimal variation 
of the field variables, requires to integrate in the path integral over a
field-configuration space which is closed under smooth deformations. The smallest 
domain satisfying this condition consists of a single homotopy class, describing
the propagation of a fixed number of kinks. The spontaneous breakdown of the
fundamental group symmetry is realized by the dynamical restriction
of the integration domain of the path integral for infinitely large systems.
The SG model is asymptotically free in the UV limit in the symmetry broken 
phase and its semiclassical solution reveals stable kinks. The important
message of this line of thought from our point of view is that the dynamical stability of
kinks lends stability for certain inhomogeneous configurations in the blocking,
\eq{block} and thereby may prevent QC to be realized.
\begin{figure}[ht]
\begin{center} 
\epsfig{file=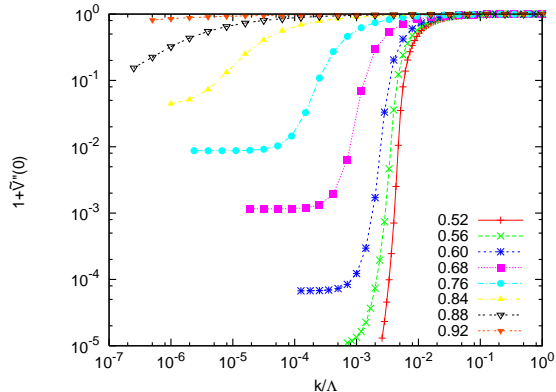,width=5.3cm,angle=-90}
\caption{\label{fig:SI}
The saturation of the curvature at small $k$ for various values of $\beta_r$.
The value of $\beta_r$ are shown for each curve.} 
\end{center} 
\end{figure}
The numerical results confirm this tendency. The scale dependence of 
the degeneracy $1+\tilde V''(0)$, depicted in Fig. \ref{fig:SI},
indicates that the loop corrections renormalize the action to a non-degenerate, scale
invariant form below a crossover scale which moves in the IR direction as we penetrate
into the symmetry broken phase. This result, not foreseen in the previous RG calculations,
requires an unconstrained treatment of the local potential.
The horizontal segments in Fig. \ref{fig:SI} signal that QC operates for 
$\beta^{(2)}_r<\beta_r<\beta^{(1)}_r$ but with a strength which decreases
with $\beta_r$. The decrease of QC forces $1+\tilde V''(0)$ to drop earlier 
during the evolution which increases the scale windows of the almost degenerate action.
For $\beta_r<\beta^{(2)}_r$ either the trajectory
is stabilized at a degeneracy level which is not detectable within the accuracy
of the double precision number representation or the steep drop of the degeneracy
continues until we hit true degeneracy and eventually generate saddle points.
\begin{figure}[ht] 
\begin{center} 
\epsfig{file=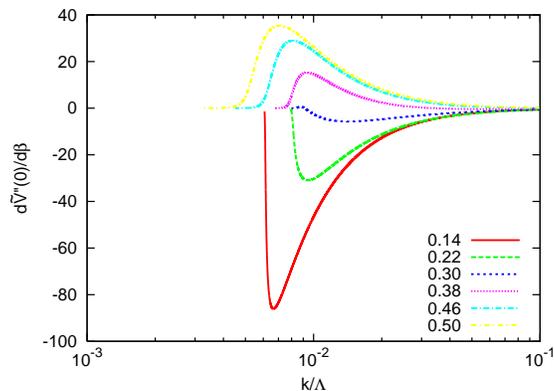,width=5.3cm,angle=-90}
\caption{\label{fig:sens}
The sensitivity matrix element up to a sign, $S_0$, as the function of the scale for 
various $\beta_r$. The value of $\beta_r$ are shown for each curve.} 
\end{center} 
\end{figure}
The degree of degeneracy of the action can be further explored by means 
of the sensitivity matrix  $\S_k$ whose elements are the derivatives of the renormalized 
quantity $\t V''_k(0)$ with respect to the bare parameters, in particular
$\S_k = \partial \t V''_k(0)/\partial \beta_r$
in the present case which is shown in Fig. \ref	{fig:sens}. For approximately $\beta_r>\beta^{(4)}$
it starts with negative values at the cutoff (not visible in the Figure)
but traverses zero and becomes positive at a 
scale which moves in the UV direction as $\beta_r$ is raised. For 
$\beta^{(3)}<\beta_r<\beta^{(2)}$ the positive peak is higher and the
negative one is lower than one. For $\beta^{(4)}<\beta_r<\beta^{(3)}$ the two peaks have 
comparable heights. There is a separatrix in the renormalization group flow
at $\beta_r\sim\beta_r^{(3)}$ reflecting a competition between Coleman and Gaussian FP, cf.
Fig. \ref{fig:phase}. Finally, for approximately $\beta_r<\beta^{(4)}$ $\S_k$ stays negative and
develops a strong peak. Such a dependence is consistent with a strengthening
degeneracy as $\beta_r$ is decreased between $\beta^{(2)}$ and $\beta^{(3)}$. 
The further decrease of $\beta_r$ seems to increase the degeneracy even more.
The dominant, IR part of the matrix element obeys the approximate scaling law 
$\S_k\approx\pm1/k^2$ with a high accuracy until the evolution changes abruptly for $\beta_r<\beta^{(4)}$.
It is far from clear if this is a precursor of a turn towards degeneracy.

Note that the appearance of the special values $\beta_r^{(n)}$ in the features discussed 
above is natural, these are the $\beta_r$ values where the UV critical exponent of a 
coupling constant changes sign, altering in a profound way the competition between the 
various Fourier modes in approaching the degenerate action.

\section{Summary} 
It is pointed out in this paper that characteristically
classical dynamics, such as the classical collective coordinate generated
Maxwell-cut, might be mimicked by quantum fluctuations with a surprising accuracy. 
Such a smearing of the usual singularities of the tree-level contributions is called QC. 

The thumb rule to estimate the strength of QC is to find the
strength of fluctuations which may have similar effects than the classical saddle
points. Note that the functional integration in question, the blocking \eq{block},
is over an UV subspace of configurations $\phi'$ only and the IR background field 
$\phi$ can stabilize inhomogeneous saddle points even if the full functional integral of 
the theory possesses no such saddle points. The fluctuations around the
saddle points are strong in general if the action has a shallow minimum at 
the saddle point. The fluctuations which may wash different saddle points together
are strong if different saddle points are close in the field space. The distinguished
feature of the SG model is the periodicity of its potential which allows us to
control the latter type of fluctuations by the parameter $\beta_r$ and makes this 
model a good testing ground for QC.

Circumstantial evidences were presented for the gradual weakening QC 
in the two-dimensional SG model as the period length of the potential is increased
in the field space. But the final word about the fate of QC in 
the small $\beta_r$ part of the phase diagram remains a provocative open problem.

We have considered vacuum expectation values in this work from the point of
view of QC. Another issue what remains open whether the full classical behavior,
decoherence included can be reproduced by QC. We plan to return to this problem
in a future publication.

S. Nagy, J. Polonyi and K. Sailer acknowledge an MTA-CNRS grant.

\end{document}